\documentclass[12pt]{article}

\usepackage{hyperref}
\usepackage{epsfig}
\usepackage{amssymb}
\usepackage{amsmath}                          
\usepackage{graphicx}
\usepackage{epstopdf}
\usepackage{setspace}

\def\simge{\mathrel{%
   \rlap{\raise 0.511ex \hbox{$>$}}{\lower 0.511ex \hbox{$\sim$}}}}

\def\simle{\mathrel{
   \rlap{\raise 0.511ex \hbox{$<$}}{\lower 0.511ex \hbox{$\sim$}}}}

\def\s#1{\setbox0=\hbox{$#1$}%
\rlap{\ifdim\wd0>.7em\kern.22\wd0\else\kern.1\wd0\fi /}#1}

\newcommand{\al}{\alpha}
\newcommand{\be}{\beta}
\newcommand{\ga}{\gamma}
\newcommand{\de}{\delta}
\newcommand{\la}{\lambda}
\newcommand{\matel}[3]{\langle #1|#2|#3\rangle}
\newcommand{\vev}[1]{\langle #1 \rangle} 
\newcommand{\state}[1]{|#1\rangle}

\newcommand{\pz}{v}
\newcommand{\ha}[2]{H_{#1#2}}
\newcommand{\am}[1]{a^{(#1)}}
\newcommand{\Hm}[1]{H^{(#1)}}
\newcommand{\mA}[1]{q_A}
\newcommand{\mB}[1]{q_B}
\newcommand{\mC}[1]{q_C}
\newcommand{\mAs}[1]{q_A^2}
\newcommand{\mBs}[1]{q_B^2}
\newcommand{\mCs}[1]{q_C^2}

\newcommand{\pA}{q_A}
\newcommand{\pB}{q_B}
\newcommand{\pC}{q_C}
\newcommand{\pI}{q_r}

\usepackage{color}
\definecolor{violet}{rgb}{0.94, 0.2, 0.8}
\definecolor{lightblue}{rgb}{0.39, 0.58, 1.00} 
\definecolor{lightgreen}{rgb}{0.1, 0.73, 0.33}

\newcommand{\gpol}{\omega}
\newcommand{\HA}{H}

\begin{document}

\begin{titlepage}
\begin{flushright}
\begin{tabular}{l}
  Edinburgh 2013-17 \\
    CP3-Origins-2013-036 DNRF90  \\
    DIAS-2013-36
\end{tabular}
\end{flushright}

\vskip1.5cm
\begin{center}
  {\Large \bf Endpoint Symmetries of Helicity Amplitudes} 
  \vskip1.3cm 
  {\sc  Roman Zwicky }
  \vskip0.5cm
  
   {\sl School of Physics and Astronomy, \\ University of Edinburgh, \\
    Edinburgh EH9 3JZ, Scotland} \\ 
  \vspace*{1.5mm}
\end{center}

\vskip0.6cm

\begin{abstract}
 We investigate helicity amplitudes (HAs)  of $A \to B C$-type decays  for arbitrary spin  
 towards the kinematic endpoint.  We show that HAs are proportional to product of Clebsch-Gordan coefficients (CGC)  and the velocity to a non-negative power. The latter can be zero in which case the HA is non-vanishing at the endpoint.
At the kinematic endpoint the explicit breaking of rotational symmetry, through the external momenta, is restored and the findings can be interpreted as a special case of  the Wigner-Eckart theorem.
Our findings are useful for i) checking theoretical computations and ii) the case 
where there is a sequence of decays, say  $B \to B_1B_2$  with the pair $(B_1B_2)$ not interacting (significantly) with the $C$-particle. 
Angular observables, which are ratios of HAs, are  given 
by ratios of CGC at the endpoint. 
We briefly discuss power corrections in the velocity to the leading order.
\end{abstract}

\end{titlepage}

\tableofcontents

\section{Introduction}

The helicity basis, as introduced by Jacob and Wick \cite{JW59}, has proven to be 
a powerful tool in describing (sequential) one-to-two decays and beyond \cite{Haber:1994pe,Richman:1984gh}.
In differential angular distributions helicity amplitudes (HAs) contain the dynamic 
information and the angular distribution is encoded in so-called Wigner $D$-functions.
In this paper we point out that at the kinematic endpoint the helicity amplitudes are simply
proportional to the Clebsch-Gordan coefficients (CGC) of  $SO(3)$ and $SU(2)$ for bosons 
and fermions in the decays.  The kinematic endpoint corresponds to the situation 
where the back-to-back velocity $v$ of the $B$ and $C$-particle in the restframe of 
$A$ is zero, i.e. all the three particles $A$, $B$ and $C$ are at rest.

Intuitively the relations can be understood as a consequence of the restoration of the spatial rotation symmetry at the kinematic endpoint. The rotational symmetry is explicitly broken by the relative momenta in the decay.
Thus the rotational symmetry  acts like a global  internal symmetry,  
analoguous to the isospin symmetry,  
and leads to  simple relations amongst (helicity) amplitudes in terms of CGC.
We wish to add that whereas the simplicity of HAs at the kinematic endpoint has been noted  in examples, e.g. \cite{Korner:1991ph}, we are not aware of 
an explicit documentation.\footnote{Our findings have been known in terms 
of the $LS$-formalism in indirect ways to specialists cf. \cite{MS}. We are grateful 
to J\"urgen  K\"orner for making us aware it.}
The extension of endpoint relations to  effective field theories is discussed in
a follow up paper \cite{HZ13}.

The paper is organised as follows: In section \ref{sec:HA} we define the HAs  
and establish endpoint degeneracies in terms of the CGC first for integer and then for half-integer 
spin in sections \ref{sec:integer} and \ref{sec:half-integer} respectively.   The asymptotic behaviour towards the endpoint and selection rules are discussed
in sections \ref{sec:asym} and \ref{sec:selection} respectively.  
The relation to the Wigner-Eckart theorem is outlined in section \ref{sec:WE}.
A total of eight examples are given
in section \ref{sec:examples} including selection rules 
in section \ref{sec:prop_examples}.
A brief summary of applications is given in section \ref{sec:applications} with particular 
focus on the isotropicity of the angular distribution of $A \to (B \to B_1 B_2) C$ at the endpoint whose derivation is deferred to appendix \ref{app:isotropic}. 
The paper ends with conclusions  section \ref{sec:conclusions}. 
Conventions on polarisation vectors and remarks on higher 
spin polarisation tensors are deferred to appendix \ref{app:pol}.

\section{Endpoint symmetries of helicity amplitudes}
\label{sec:HA}

To begin with we consider the $ 1 \to 2$ decay of particles with generic spin
\begin{equation}
\label{eq:ABC}
A(\pA , \la_A) \to B(\pB, \la_B)   C(\pC, \la_C) \; ,
\end{equation}
where $\pA = \pB+\pC$ are the corresponding momenta and $\la_r$ ($r \in \{A,B,C\}$ hereafter) denote the helicities. Angular momentum conservation imposes 
\begin{equation}
\la_A = \la_B -\la_C = \la_B + \bar \la_C \;,
\end{equation}
 ($\bar \la \equiv -\la$ hereafter).
The HAs are,
\begin{equation}
\ha{\la_B}{\la_C} = {\cal A}\left( A(\la_A) \to B(\la_B) C(\la_C) \right) \;,
\end{equation}
the amplitudes for fixed helicities of the $B$- and $C$-particle.

In general there are many Lorentz-invariant structures of the amplitudes 
\begin{equation}
\label{eq:dec}
\ha{\la_B}{\la_C} = \sum_{i=1}^n a_i( \pA^2  ,\pB^2, \pC^2 ) P_i(\al,\be^*(\la_B),\ga^*(\la_C),\pB,\pC)  \;,
\end{equation}
where $\alpha$, $\beta$ and $\gamma$ (indices suppressed) are the corresponding polarisation tensors and $n$ is usually a number well below ten. 
We refer to the coefficients $a_i$ as form factors.
In the case where any of the  particles $A$, $B$ or $C$ are off-shell $\pI^2 \neq m_r^2$, such as in further sequential decay $B \to B_1 B_2$, the propagation of the particle $i$ 
can be approximated by a Breit-Wigner ansatz.
The symbols $P_i$ denote Lorentz-invariant structures built out of the momentum vectors $\pB,\pC$ ($\pA = \pB+\pC$ is dependent) and the corresponding  polarisation tensors  $\al(\pA,\la_A),\be(\pB,\la_B),\ga(\pC,\la_C)$. 
The invariants $P_i$ are linear in the polarisation tensors.
 
 Before following the approach outlined in \eqref{eq:dec} 
we motivate or demonstrate the findings of this papers in the language of the
symmetry breaking and the Wigner-Eckart theorem.
 The global Lorentz-symmetry is broken by the momenta of the particles $A,B$ and $C$,  but at the  kinematic endpoint,\footnote{Since  particle $C$ is moving into the opposite direction one has to flip the momenta. 
Intuitively it is clear that this leads to a helicity flip and thus one has to use 
 $\bar \la_C$ instead of $\la_C$. As usual in quantum theory there can  be a helicity dependent phase  associated with this transformation. 
 The important point is that  in the Jacob-Wick convention, which we adapt throughout, 
  this phase is conveniently set to unity.
Some more detail is given in appendix \ref{app:pol} and in particular section \ref{app:second}.},
\begin{equation}
\label{eq:momenta}
\pA \propto \pB \propto \pC \propto \gpol(t) \equiv (1,0,0,0) \;, 
\end{equation}
where the momentum 
vectors are all proportional to the time direction 
the vectorial $SO(3)$ and spinorial $SU(2)$  global symmetry is restored.

Essentially we are left with the problem of combining $SU(2)$ representation 
into an invariant.  The latter is  non-vanishing if and only if 
the spin of the three particles can be added to zero which is the case if 
\begin{equation}
\label{eq:cond}
|J_B-J_C| \leq J_A \leq J_B+J_C\;.
\end{equation}
In turn of the more familiar  Kronecker product this reads,
 \begin{equation}
 \label{eq:ABC}
 \left(  \mathbf{J_A}   \otimes   \mathbf{J_B}  \otimes  \mathbf{ J_C} \right)_{SU(2)} = 1 \cdot \, \mathbf{0} \oplus .. \;,
 \end{equation}
 where $\mathbf{J}$ stands for the $J$-spin representation throughout 
 and $ \mathbf{0}$ is the trivial representation in this notation. 
 An important point is that $SU(2)$  is \emph{simply reducible} which means that 
 the multiplicity of each irreducible representation is one when the Kronecker product 
 of two  irreducible representations is taken and this justify the $1$ in front of the 
 \eqref{eq:ABC}. In practice this implies that there's a \emph{unique} (leading)  HA at the
 kinematic endpoint.
 \footnote{Excluding the case where both 
 decays $A \to B C$ and $B \to B_1 B_2$ are parity-violating this leads to an 
 angular distribution at the endpoint which is free from hadronic parameters. 
 A rare counterexample to this case is $\Lambda_b \to \Lambda(\to N \pi) \ell \nu$ 
 where $\Lambda\to N \pi$ decays via the weak force since it is stable under QCD 
 \cite{HZ21}.}
 
Denoting the state of total angular momentum $J$ and helicity 
$\lambda$ by $\state{J,\lambda}$, as usual, the HA is proportional to the 
scalar product of 
$\state{J_A,\la_B+\bar \la_C}$  with the state $ \state{J_B, J_C ; \la_B \bar \la_C  } \equiv  \state{J_B; \la_B }\otimes \state{ J_C ;  \bar \la_C  }  $. 
This is precisely the definition of the CGC\footnote{Throughout this paper we are assuming the  Condon-Shortly convention (c.f  appendix \ref{app:pol}) for 
which the CGC are real. The values of CGC can for instance be looked up  in the Particle Data Group Book \cite{PDG}.},
\begin{equation}
C^{J  J_1 J_2}_{(\la_1 +  \la_2) \la_1  \la_2} \equiv \vev{ J_1, J_2 ; \la_1  \la_2     | J,\la_1+ \la_2}    \;.
\end{equation}
 The HA at the endpoint is therefore given by 
\begin{equation}
\label{eq:geometry}
\ha{\la_B}{\la_C}  =   \am{0} \,  C^{J_A J_B J_C}_{(\la_B +  \bar \la_C) \la_B \bar \la_C}  \;,   
\end{equation} 
where $\am{0}$ is proportional to the single form factor at the endpoint.   
N.B. $\am{0}$ generally is a linear combination out of the set  $\{a_1,a_2, ..\}$ outlined in Eq.~\eqref{eq:dec}. 
In the following section \ref{sec:examples} we will discuss the case when it is proportional to a power of the velocities and selection rules through examples showing the working of Eq.~\eqref{eq:geometry}.
We will  demonstrate the specifics on examples in section \ref{sec:examples}.

We wish to emphasise that the other combinations of form factors do not vanish 
as matrix elements per se, but their respective Lorentz structures $P_i$ do.
We comment in section \ref{sec:WE} on the relation to the Wigner-Eckart theorem 
of \eqref{eq:geometry}.
One immediate consequence of \eqref{eq:geometry} is that (for non-vanishing HAs the endpoint)
the uniangular distribution, in the angle 
$\theta_B$ between $B_1$ and the $B$-particle in $A \to (B \to B_1 B_2) C$,
 is isotropic in the angle if one sums over initial state polarisation  and not both 
 decays are parity violating 
 c.f. appendix \ref{app:isotropic}.
Intuitively this corresponds to the situation where the particles
have lost any spatial reference point and can therefore not decay into a particular direction 
more frequently than in any other.

\subsection{The invariants}

In the two following subsection we discuss how the unique invariant in \eqref{eq:ABC}
works out in the case of integer and half-integer spin in sections \ref{sec:integer} 
and \ref{sec:half-integer} respectively.

\subsubsection{Integer spin}
\label{sec:integer}

In the case of bosons 
the spatial symmetry part of the Lorentz group corresponds to $SO(3)$.
The defining, i.e. invariant, tensors in the fundamental representation of $SO(3,1)$ are
the metric $g_{\mu\nu}$ or the  totally antisymmetric 
Levi-Cevita tensor (LCT) $\epsilon_{\mu\nu\rho\sigma}$. The former corresponds to $O(3,1)$ invariance and the latter ensures unit determinant of the Lorentz transformation.
In the following analysis we will see that these tensors effectively reduce to the $SO(3)$ 
defining tensors.

To establish this claim we note that:
i) At the endpoint the $J=1$ polarisation vectors, denoted by a hat, 
are all proportional to each other 
$\gpol(\la) \equiv \hat \al(\la) = \hat \be(\la) =  \hat \ga(\bar \la)$ (c.f. appendix \ref{app:pol}).  
N.B. the barred polarisation  index  ($\bar \la_C  \equiv -\la_C$) is used for convenience.
ii) The polarisation tensors 
$\al_{\mu_1 ..\mu_{J_A}}(\pA,\la_A), \be_{\mu_1 ..\mu_{J_B}}(\pB,\la_B)$ and 
$\ga_{\mu_1 ..\mu_{J_C}}(\pC,\la_C)$ can be formed out of the $J=1$ polarisation 
vector through appropriate CGC as described in appendix \ref{app:higher}. 
Armed with this knowledge the following two facts are of importance to establish the claim:
\begin{itemize}
\item[(a)] \emph{Transversity:} 
Any contraction of $\gpol(t)$ with $\al,\be,\ga$ is zero \eqref{eq:transverse} since $\gpol(t)$ is, so to speak, the direction 
of the momentum \eqref{eq:momenta}. 
\item[(b)] \emph{Anti-symmetric tensor:} The only object with which $\gpol(t)$ can  be contracted without vanishing is the LCT to 
$\epsilon_{mno} \equiv \epsilon_{mno\rho}\gpol^{\rho}(t)$.  Furthermore, since any product of LCT can expressed in terms
of metric tensors it is sufficient to consider a single LCT.
\end{itemize}
By virtue of (a) one can 
safely replace the  metric $g_{\mu\nu}  \to -\delta_{mn}$, where $\delta_{mn}$ denotes the Kronecker symbol throughout this work. Roman indices run from $1,2,3$ (spatial indices) and Greek indices run from 
$0,1,2,3$ (spacial and temporal indices).  The tensors $\delta_{mn}$ and $\epsilon_{mno}$ are the defining tensors of $SO(3)$ 
and we have thus justified our initial claim at the beginning of this section.

 \subsubsection{Extension to half-integer spin}
 \label{sec:half-integer}

The new element with respect  to integer spin is that one can form 
covariants out of two half-integer spin objects. 
Since $(n+1/2)$-spinors can be formed out of integer spin and $1/2$-spinors(cf. 
 section \ref{sec:32}), we may restrict our attention to $1/2$-spinors.  
The important objects are the particle and anti-particle spinors $u$ and $v$.
In the Dirac representation of the Clifford algebra, with $\sigma^i$ as the usual $2\times 2$ Pauli matrices,
\begin{equation}
\label{eq:Dirac}
 \ga^0 = \begin{pmatrix} 1&0 \\ 0& -1 \end{pmatrix}\,, \qquad
  \ga^i = \begin{pmatrix} 0 & \sigma^i\\ -\sigma^i & 0 \end{pmatrix}\,, \qquad
  \ga_{5} = 
\begin{pmatrix} 0& 1\\ 1& 0 \end{pmatrix}  \;,
\end{equation}
 $u$ and $v$ (e.g. \cite{Haber:1994pe}) assume a simple form at the endpoint:
 \begin{equation}
 \label{eq:uv}
u(\vec{p} = 0,\la) = \begin{pmatrix} \chi_\la \\ 0 \end{pmatrix}  \;, \quad  v(\vec{p} = 0,\la) = 
\begin{pmatrix}  0 \\  -2\la \chi_{\bar \la} \end{pmatrix}  \;. 
\end{equation}
 The symbol $\chi_\la$ denotes a  $2$-spinor which does not need to be specified any further 
 for our purposes. To proceed further we need to determine the selection rules 
 for the spinor products of the form $L_{uu}^{\Gamma} =  \bar u \Gamma u$  ($f_1 \to f_2 b$) and 
 $L_{vu}^{\Gamma} = \bar v \Gamma u$ ($b \to \bar f_1 f_2 $) where $b$ and $f$ stand for boson and fermion 
 respectively, and $\Gamma$ is a combination of the Dirac matrices. For the corresponding 
 anti-particle decay one has to exchange $u$ and $v$ with each other.
 The sets  $\Gamma_D = \{\ga_0, \mathbf{1} \}$ and $\Gamma_A =
 \{ \ga_i,\ga_5 \}$ induce a grading in the sense that any product of an even 
 number of $\Gamma_A$-matrices is in the $\Gamma_D$ set. 
Directly relevant to us is  that  either $L_{uu}^{\Gamma_D}$ or $L_{uu}^{\Gamma_A}$ are
zero at the endpoint and the same is true for all other combinations of $u$ and $v$.
This effectively reduces the $SO(3,1)$ symmetry to $SO(3)$ as $\Gamma_{A,D}$ discriminate between 
 spacial and temporial indices.
 
 Let us illustrate this statement with one example. The decay of 
 spin $1$ boson into two spin 1/2 fermions.  The Lorentz invariant is given by 
 \begin{equation}
 \bar v(\pB,\la_B) \ga_\mu  u(\pC,\la_C) \, \alpha_\nu(\pA,\la_A) g^{\mu\nu} \,.
 \end{equation}
  Eqns (\ref{eq:Dirac},\ref{eq:uv}), allows to replace $g_{\mu\nu} \to -\delta_{mn}$,
 making the $SO(3)$-symmetry explicit.  Note, for example the replacement 
 of $\ga_\mu \to \ga_\mu \ga_\nu \pI^\nu$ would not bring in anything new since at the endpoint 
 the latter reduces to  $\ga_\mu \ga_0 \pI$ and differs therefore by a constant only. Furthermore,  we  
 see that $\ga_\mu \to \ga_\mu \ga_5$ vanishes since $\ga_5$ is in the other class  
(cf.  rule (iii) in section \ref{sec:selection}).
 The case where there is half-integer spin in the initial and final state is analoguous with the crucial difference that the other class has to be chosen as now the product 
 $\bar u .. u ( \bar v .. v)$ has to be investigated.
 
 In conclusion, even when there are fermions amongst the 
 particles $A$, $B$ and $C$, 
 there is a unique form factor that enters at the kinematic endpoint and 
 formula \eqref{eq:geometry} remains correct for fermions in the decay as well.
 The discussion in this paragraph does have formal reminiscence 
with heavy quark effective theory. Although we wish to stress that the
latter is a dynamic theory and that our situation is of merely kinematic nature.

\subsection{Helicity amplitudes $\propto  \pz^\Omega$ towards the endpoint}
\label{sec:asym}

The goal of this section is to generalise \eqref{eq:geometry} 
to the case when  the condition \eqref{eq:cond} is not met.
In order to do so it us useful to define $\Omega$: 
 the number of open polarisation indices of $\al,\be,\ga$ after maximal contractions. 
Formally it may be written as:
\begin{equation}
\label{eq:Om}
\Omega  = \Omega(J_A;J_B,J_C) =  \begin{cases}
0  & |J_B-J_C| \leq J_A \leq J_B+J_C  \\[0.1cm]
p & \text{otherwise}\end{cases} \;,
\end{equation}
with $p \equiv \min\{ |J_A-  |J_B-J_C|| ,|J_A - (J_B+J_C)| \}$.
For example if $(J_A;J_B,J_C) = (4;1,1)$ then $\Omega =2$. 
If $\Omega > 0$ then necessarily all the open indices are in one polarisation 
tensor,  say $\al_{\mu_1 ..\mu_\Omega}$, as otherwise one would just contract 
the indices of different polarisation tensors. Furthermore, the indices of $\al_{\mu_1 ..\mu_\Omega}$ 
cannot be contracted any further by the LCT since the polarisation tensors 
are totally symmetric in the indices, and neither with the metric since they are traceless \eqref{eq:traceless}.
The open polarisation indices of $\al$  have to be contracted by the momenta $\pB$ 
which differs linearly from $\pz$ in  $\pA$. Therefore the asymptotic behaviour is given by
\begin{equation}
\label{eq:scaling}
\ha{\la_B}{\la_C} = \,  \Hm{\Omega}_{\la_B \la_C} \,  \pz^{\Omega}  + ..\;,
\end{equation} 
where the dots stand for higher order corrections in $v$.
Eq.~\eqref{eq:geometry} corresponds to $\Omega = 0$. For $\Omega > 0$ 
one can get $\Hm{\Omega}_{\la_B \la_C}$  as follows.  Let $J_A > J_{B,C}$ be the largest of all three spins. 
It is useful to think of a fourth fictitious  particle, sometimes called spurion in other contexts,  of spin $\Omega$ with helicity $0$.  And then consider 
the following two Kronecker products:
\begin{alignat}{2}
& (\mathbf{J_B} \otimes \mathbf{J_C})_{SU(2)} &\;=\;& 1 \cdot (\mathbf{J_B+J_C}) \oplus ..  \;, \nonumber  \\[0.1cm]
& (\mathbf{J_A} \otimes \mathbf{\Omega})_{SU(2)} &\;=\;&1 \cdot (\mathbf{J_B+J_C}) 
\oplus .. \;  \;,
\end{alignat}
 which in turn can be combined into a singlet. 
Above the dots stands for other representations that are not of interest here. 
 We get
\begin{equation}
\label{eq:scaling2}
 \Hm{\Omega}_{\la_B \la_C} = \begin{cases}
\am{0} \,  C^{J_A J_B J_C}_{\la_A \la_B \bar \la_C}    &\Omega = 0  \\[0.1cm]
\am{\Omega} \, C^{(J_B+J_C) J_B J_C}_{\la_A \la_B  \bar \la_C }
C^{(J_B+J_C) J_A \Omega}_{ \la_A \la_A 0} & \Omega > 0 \end{cases} \;,
\end{equation}
where we have used  $\la_A = \la_B + \bar \la_C$ for compact notation.
We note that the formula for $\Omega > 0$ reduces to $\Omega =0$ since 
$C^{J_AJ_A0}_{  \la_A \la_A 0} = 1$. The relation to the Wigner-Eckart theorem 
is discussed in section \ref{sec:WE} and constitutes an effective proof of the formulae as opposed to the pedestrian analysis given here. In particular it also constitutes 
the proof for when there are fermions in the decay.

Whether or not $\am{\Omega}$ vanishes for  reasons other than  spin is discussed in  
section \ref{sec:selection}  and illustrated in section \ref{sec:prop_examples} with examples.

\subsubsection{Comments on power  corrections}
\label{sec:NLO}

The 
 notation of \eqref{eq:scaling}  is extended to include the first correction (where $n > 0$ and integer)  as follows,
\begin{equation}
\label{eq:scalingNLO}
\ha{\la_B}{\la_C} = \,  \Hm{\Omega}_{\la_B \la_C} \,  \pz^{\Omega}  +   \Hm{\Omega+n}_{\la_B \la_C} \,  \pz^{\Omega+n}  + {\cal O}(v^{\Omega+n+1})\;,
\end{equation} 
and we refer to such corrections as N$^n$LO power corrections.  
Note that in the case of \emph{parity conservation}  
(cf. rule (ii)  in section \ref{sec:selection}),  $n$ is  \emph{even}. 
This is of relevance as only NLO corrections ($n=1$) to the asymptotic behaviour can   be understood in terms of CGCs \eqref{eq:scaling2}.  
However,  these corrections are not necessarily governed by a unique 
invariant. If one of the three particles happens to be of spin $0$ then the invariant is 
still unique such as in $B \to K^* (\ga^* \to \ell^+ \ell^-)$ for example. 
If there are two $k$ invariant at NLO then the differential distribution 
in the vicinity of the endpoint will be governed by $k$ ratio of form factors. 
For examples cf. section \ref{sec:NLO-example} and  footnote \ref{foot:two} in that section.

\subsection{Selection rules beyond spin}
\label{sec:selection}

In this section we discuss selection rules beyond the spin condition \eqref{eq:cond} 
that apply to the  $\am{0}$ and  $\am{\Omega} \pz^\Omega$-terms.

In the case where parity is conserved the following constraint applies to 
the HA (e.g.  \cite{JW59,Haber:1994pe}):
\begin{equation}
\label{eq:inversion}
 \ha{\la_B}{\la_C} =  \eta \, (-1)^{\Delta J}  \ha{\bar \la_B}{\bar \la_C}  \;,
\end{equation}
where 
\begin{equation}
\label{eq:DJeta}
\Delta J\equiv  (J_B+J_C)-J_A \;, \quad    \eta \equiv \eta_A \eta_B \eta_C \;, 
\end{equation}
and $\eta$ is the product of  the intrinsic  parities  of the particles $A$, $B$ and
$C$. This has to be put into context with the CGC-symmetry property \cite{Hecht},
\begin{equation}
\label{eq:CGCbar}
C^{J_A J_B J_C}_{\la_A \la_B \bar \la_C}   = 
(-1)^{\Delta J}  C^{J_A J_B J_C}_{\bar \la_A \bar \la_B   \la_C}   \;,
\end{equation}
and Eq.~\eqref{eq:geometry}.
We note that the following rules must apply:
\begin{itemize}
\item \emph{if parity is conserved} 
\begin{itemize}
\item[(i)]
The total internal parity  must be one, $\eta =1  $, for the HA not to vanish at the endpoint 
for \eqref{eq:geometry}, \eqref{eq:inversion} and \eqref{eq:CGCbar} to be consistent with each other.
\item[(ii)] In fact (i) is implicit in a more general theorem (e.g. \cite{Weinberg:1995mt})
 that states  
that $\eta$ equal to $1(-1)$ implies that the  S-matrix is even (odd) in powers of the external momenta. Let us write the statement in mnemonic form:
\begin{equation}
\eta =  \begin{cases}
+1  & $S$ \text{ even external momenta }  \\[0.1cm]
-1  & $S$ \text{ odd external momenta } \end{cases} \;.
\end{equation}
A non-vanishing HA at the endpoint is constant in $\pz$ and therefore an even power 
implying $\eta =1$ as in (i).
\end{itemize}

\item \emph{if parity is not conserved} 
\begin{itemize}

\item[(iii)]  It is easy to convince oneself, using arguments along the lines of
section \ref{sec:half-integer},  that for  
odd and even powers of $\Delta J + \Omega$ (with 
$\Omega,\Delta J \equiv (J_B+J_C) - J_A$ as in \eqref{eq:Om} and \eqref{eq:DJeta})
there is a definite association with the endpoint amplitude  
and the tensorial or spinorial structure. 
We use the following cryptic notation: $(\epsilon)$ and $(g)$ depending on whether 
the tensors contain a  LCT or not.  $(\ga_5)$ or $(\mathbf{1})$ depending on whether 
the Dirac spinor product contains a $\ga_5$ or not.
With the notation (\ref{eq:scaling},\ref{eq:scaling2}), 
$ \text{HA} \propto a^{(\Omega)} \pz^\Omega $, the result can be written in mnemonic form as follows:
\begin{eqnarray}
\label{eq:DeltaJ}
 \Delta J + \Omega =  \begin{cases}
\text{odd}    &  a^{(\Omega)} \text{ contains: } \epsilon \text{ (bosons), }   \ga_5\text{ (fermions)}  \\[0.1cm]  
 \text{even} & a^{(\Omega)} \text{ contains: } g \text{ (bosons), }  \mathbf{1}\text{ (fermions)}   \end{cases} 
\end{eqnarray}
\end{itemize}
\item \emph{Identical particles (Landau-Yang-type selection rule)}
For identical particles the following relation holds \cite{JW59}:
$\ha{\la_B}{\la_C} =  (-1)^{\Delta J}  \ha{ \la_C}{\la_B}  $. 
\begin{itemize}
\item[(iv)]
The above formula implies that 
for $\Delta J$ odd $\HA_{\la\la} = 0$ .  If this is the case then 
 Eq.~\eqref{eq:geometry} implies that all amplitudes have to vanish at the endpoint. More precisely the form factor $\am{0}$ has to vanish since not all CGC are zero. 
 An example is given in section \ref{sec:ZZ} for $1 \to 1+ 1$ decay.
 \end{itemize}
\end{itemize}

\subsection{Relation to the Wigner-Eckart theorem}
\label{sec:WE}

We  briefly  comment on the relation of the formulae \eqref{eq:scaling2} to the Wigner-Eckart theorem.
The latter states that (e.g. \cite{Hecht}),
\begin{equation}
\label{eq:WE}
\matel{jm}{T^q_k}{j'm'} = C^{jkj'}_{mqm'} \matel{j}{|T^q|}{j'}
\end{equation}
the matrix element of a tensor operator is determined by a product of CGC and a reduced matrix element which is independent on the orientation (helicity).
First we discuss the somewhat degenerate $\Omega = 0$-case. 
Note that the CGC, $C^{J_A J_B J_C}_{\la_A \la_B \bar \la_C}$ in \eqref{eq:scaling2}, is  the basis transformation 
from the state $\state{J_B, J_C ; \la_B \bar \la_C  } $ to $\state{J_A,\la_B+\bar \la_C}$.
The form factor  $\am{0}$ takes the role of  the reduced matrix element, independent of
the helicities, where the transition operator is a scalar (namely the Hamiltonian of the decay).
The $\Omega > 0$-case is more interesting as it can be viewed as the reduced matrix element 
of a tensor operator with angular momentum $q= \Omega$ and helicity $k =0$  which is then contracted by an appropriate momentum  $\pI$. Thus the $C^{(J_B+J_C) J_A \Omega}_{ \la_A \la_A 0}$ in the second line of \eqref{eq:scaling2} corresponds to the CGC in \eqref{eq:WE}.

\section{Eight Examples}
\label{sec:examples}

We  illustrate our main formulae \eqref{eq:geometry} and \eqref{eq:scaling2} 
by a few examples found 
in the literature as well as  an example that we work out explicitly. 
In this section we frequently  use the short hand notation $J_A^{P_A} \to J_B^{P_B} + J_C^{P_C}$ 
to indicate the spin ($J_i$) and the parity ($P_i$, if known) of the particles involved. 
In the tables given below we have not made  use of the symmetry property \eqref{eq:inversion}
but have listed all the values explicitly.

\subsection{Higgs-like (spin $0,1,2$) decay into two $Z$-bosons}
\label{sec:ZZ}
We consider  $H_{J=0,1,2} \to Z^* Z$ where the particle $H$ has got spins $0,1$ or $2$. 
By Higgs-like we mean that we are open to other spin than zero for the decaying particle.
We are going to use the result in Ref.~\cite{Hopkins12} for which the endpoint is 
given by  $m_X \to m_1+m_2$  (which in their notation implies $x \equiv \left[((m_X^2-m_1^2-m_2^2)/(2 m_1 m_2))^2 -1 \right] \to 0$). The variables $m_X$, $m_1$ and $m_2$ denote the masses of the Higgs-like particle, the first and the second $Z$ boson respectively. The variable $x$
 is proportional to the K\"all\'en-function (whose square root is proportional to the back-to-back velocity $v$ 
 in the rest-frame of the decaying particle) and
 is therefore zero at the endpoint.   
 
 Note that in all the examples above $\Omega =0$ so the HA, modulo question of parity, do not vanish at the endpoint. It is though immediate that for $J_A = J_H > 2$, $\Omega = J_H -2$ and thus by virtue of \eqref{eq:scaling} the 
 $\text{HA} \propto v^{(J_H-2)}\;\;$.\footnote{This special case was noted in Ref. \cite{Milleretal}. Our work adds
 the exact form \eqref{eq:scaling2} of the helicity dependent precoefficient.}

 \paragraph{$\mathbf{0 \to 1+1:}$} 
 The prediction \eqref{eq:geometry}, $\ha{\la_{B}}{\la_{B}} = a^{(0)} C^{011}_{0\la_B \bar \la_B}$,  yields: 
 \begin{center}
\begin{tabular}{l  | ccc}
HA & $\ha{+}{+}$&$\ha{0}{0}$& $\ha{-}{-}$  \\[0.1cm] \hline 
$\la_A$ & $0$ & $0$ & $0$  \\[0.1cm]
CGC &  $C^{011}_{01\bar 1}$ &$C^{011}_{000}$& $C^{011}_{0\bar 11}$ \\[0.1cm]
value &  $\sqrt{1/3}$ & -$\sqrt{1/3}$ & $\sqrt{1/3}$
\end{tabular}
\end{center}  
where we have indicated the helicity of particle $A$, the CGC and its value on lines 
one, two and three respectively. 
The table above is consistent with Eq.14  \cite{Hopkins12},  $m_X \to m_1+m_2$  ($x \to 0$) and  the identification 
$a^{(0)} = \sqrt{3} (m_1+m_2)^2/v_H \, a_1$, denoting the Higgs vacuum expectation value by $v_H$ to avoid confusion with the velocity $v$.

\paragraph{$\mathbf{1 \to 1+1:}$} 
The prediction  \eqref{eq:geometry}, $H_{\la_{B}  \la_{C}} = a^{(0)} C^{111}_{(\la_B + \bar \la_C) \la_B \bar \la_C}$, yields:
\begin{center}
\begin{tabular}{l  | ccccccc}
HA &  $\ha{+}{0}$&$\ha{0}{-}$&$\ha{+}{+}$&$\ha{0}{0}$&$\ha{-}{-}$&$\ha{0}{+}$&$\ha{-}{0}$ \\[0.1cm] \hline
$\la_A$ & $1$ & $1$ & $0$ & $0$ & $0$ & $-1$ & $-1$  \\[0.1cm]
 CGC &  $C^{111}_{110}$&$C^{111}_{101}$  & $C^{111}_{01\bar 1}$&$C^{111}_{000}$ &$C^{111}_{0\bar 11}$ & $C^{111}_{\bar 10\bar 1}$ & $C^{111}_{\bar 1\bar 10}$  \\[0.1cm]
value &  -$\sqrt{1/2}$ & $\sqrt{1/2}$  & $\sqrt{1/2}$& $0$ & -$\sqrt{1/2}$ & -$\sqrt{1/2}$ & $\sqrt{1/2}$ 
\end{tabular}
\end{center}
which agrees with Eq.17 of \cite{Hopkins12} for  $m_X \to m_1+m_2$  ($x \to 0$) and $a^{(0)} = \sqrt{2} i (m_1-m_2)$. In  the case where $m_1 =m_2$ the two $Z$ bosons are identical and the Landau-Yang theorem (rule (iv) in section \ref{sec:selection}), applies and implies $H_{\la\la} =0$. 
Hence by virtue of \eqref{eq:geometry} all HAs vanish at the endpoint. 
This is indeed the case as $a^{(0)} \propto m_1-m_2 \stackrel{m_1 \to m_2}{\to} 0$.

\paragraph{$\mathbf{2 \to 1+1:}$}
The prediction  \eqref{eq:geometry}, $\ha{\la_{B}}{  \la_{C}} = a^{(0)} C^{211}_{(\la_B + \bar \la_C) \la_B \bar \la_C}$, yields:
\begin{center}
\begin{tabular}{l | ccccccccc}
 HA & $\ha_{+-}$&$\ha{+}{0}$&$\ha{0}{-}$&$\ha{+}{+}$&$\ha{0}{0}$&$\ha{-}{-}$&$\ha{0}{+}$&$\ha{-}{0}$&$\ha{-}{+}$   \\[0.1cm] \hline
$\la_A$ &  $2$  & $1$ & $1$ & $0$ & $0$ & $0$ & $-1$ & $-1$ & $-2$   \\[0.1cm]
CGC &  $C^{211}_{21\bar 1}$&$C^{211}_{110}$&$C^{211}_{10\bar 1}$&$C^{211}_{01\bar 1}$&$C^{211}_{000}$&$C^{211}_{0\bar 11}$&$C^{211}_{\bar 1 01}$&$C^{211}_{\bar 1\bar 10}$&$C^{211}_{\bar 2\bar 11}$ \\[0.1cm]
value & $1$&$\sqrt{1/2}$&$\sqrt{1/2}$&$\sqrt{1/6}$&$\sqrt{2/3}$&$\sqrt{1/6}$&$\sqrt{1/2}$&$\sqrt{1/2}$&$1$
\end{tabular}
\end{center}
which is consistent with Eq.21 in \cite{Hopkins12} in the limit  $m_X \to m_1+m_2$  ($x \to 0$) and 
$a^{(0)} \equiv c_1 m_1 m_2/\Lambda$.  The corresponding NLO power correction is worked out in 
 section \ref{sec:NLO-example}.

\subsection{Higgs-like (spin 0,1,2) decay into a fermion pair}
\label{sec:ff}
We consider  $H_{J=0,1,2} \to q \bar q$ where the particle $H$ has got spin $0,1$ or $2$.
We are using the results of Ref.~\cite{Hopkins10} for which at the endpoint $m_X \to 2 m_q$ (and the fermion velocity $\beta = 0$) where $m_X$ and $m_q$ are the Higgs-like particle and the fermion mass respectively.
We disagree in signs with \cite{Hopkins10} for some of the $2,1 \to 1/2 + 1/2$ amplitudes\footnote{The sign differences are of no consequence for the work in \cite{Hopkins10}, because of the (incoherent) factorisation of the production helicity amplitudes and the parton distribution functions. 
In essence only  the absolute values, denoted by $B_{\la_1\la_2}$ in \cite{Hopkins10}, of the production amplitudes enter.}.

\paragraph{$\mathbf{0 \to 1/2^+ +1/2^-:}$}
One gets using \eqref{eq:geometry} yields:
$H_{\la_B \la_C} =  \,a^{(0)}  C^{0 \frac12 \frac12}_{(\la_B  + \bar \la_C)  \la_B  \bar \la_C } \;$.
Since $C^{0 \frac12 {\frac12}}_{0  \frac12 \bar{ \frac12} } = C^{0 \frac12 \frac12}_{0  \bar{ \frac12} \frac12 }$ one gets $H_{\frac12 \frac12} = H_{\bar{\frac12} \bar{\frac12}}$ 
which is in accord with \cite{Hopkins10} Eq.20 with $a^{(0)} = m_q/v_H m_X$.
\paragraph{$\mathbf{1 \to  1/2^+ +1/2^-:}$}
For spin 1 \eqref{eq:geometry} yields:
$H_{\la_B \la_C} = \,a^{(0)} C^{1 \frac12 \frac12}_{(\la_B  + \bar \la_C)  \la_B  \bar \la_C }  $.
\begin{center}
\begin{tabular}{l | cccc}
HA &  $\ha{ \frac12 }{\bar{ \frac12 }}$&  $\ha{ \frac12 }{ \frac12 }$ & 
 $\ha{\bar{ \frac12 }}{\bar{ \frac12 }}$&$\ha{\bar{ \frac12 }}{ \frac12 }$ \\[0.1cm] \hline
$\la_A$ & $1$ & $0$ & $0$ & $-1$  \\[0.1cm]
CGC &   $C^{1 \frac12 \frac12}_{1 \frac12  \frac12  }  $&$C^{1 \frac12 \frac12}_{0 \frac12  \bar{\frac12}   }  $ &$C^{1 \frac12 \frac12}_{0  \bar{\frac12}  \frac12    }  $& $C^{1 \frac12 \frac12}_{\bar 1  \bar{\frac12}  \bar{\frac12}   }  $  \\[0.3cm]
value &  $1$&  $1/\sqrt{2}$ &  $1/\sqrt{2}$ &  $1$ \end{tabular}
\end{center}
this agrees with Eq.21 \cite{Hopkins12} only up to  signs ($|a^{(0)}| = 2 m_q \rho_1^{(1)}$). 
Note though that the endpoint-relation \eqref{eq:inversion}  (with $\eta = 1$, $\Delta J =0$) reads 
$H_{\la_1 \la_2} = H_{\bar \la_1 \bar \la_2}$ and  is consistent with our results. 
As the statement has some degree of circularity,  we have also explicitly checked our results and find agreement with \eqref{eq:geometry}.
\paragraph{$\mathbf{2 \to  1/2^+ +1/2^-:}$}
In this example $\Omega = 1$ and this is a good test of the formula on 
the second line in Eq.\eqref{eq:scaling2}: 
\begin{equation}
H_{\la_B \la_C} = a^{(1)} \, C^{1 \frac12 \frac12}_{\la_A \la_B  \bar \la_C }
C^{1 2 1}_{ \la_A \la_A 0} \, \pz \;.
\end{equation}
Evaluating we get:
\begin{equation}
\label{eq:2ff}
\ha{\frac12}{\frac12}  = \ha{\bar{\frac12}}{\bar{\frac12}} =  ( \sqrt{2/3}) a^{(1)}\, \pz \;, \quad 
\ha{\frac12}{\bar{\frac12}}  = \ha{\bar{\frac12}}{\frac12}  =  a^{(1)}\, \pz \;, 
\end{equation}
which again agrees with Eq.(22) in \cite{Hopkins10} up to signs ($|a^{(1)}| = \rho_1^{(2)} m_X^2/\Lambda$.)  In the same vein linearity in $\pz$ by rule (ii) in section \ref{sec:selection} implies $\eta = -1$ and since $\Delta J = 1$ 
\eqref{eq:inversion} would suggest that $H_{\la_1 \la_2} = H_{\bar \la_1 \bar \la_2}$ which 
is consistent with \eqref{eq:2ff}.

\subsection{$\Lambda_b \to \Lambda_c(W \to \ell \nu)$ or  $1/2^+ \to 1/2^+  +   1$} 
\label{sec:12}
Formula \eqref{eq:geometry} predicts:
\begin{equation}
\frac{H_{\frac12 1}}{H_{\frac12 0}} 
= \frac{C^{\frac12 \,  \frac12 \, 1}_{\bar{\frac12} \, \frac12 \, \bar 1}}{C^{\frac12  \, \frac12 \,1}_{\frac12 \, \frac12 \, 0}}  
 = \frac{\sqrt{2/3}}{-\sqrt{1/3}} = - \sqrt{2} \;,
\end{equation}
which is the result obtained in \cite{Korner:1991ph} in Eq 4.  It is found that the axial ($J^P = 1^+$) but not the vectorial coupling is non-vanishing at the endpoint  \cite{Korner:1991ph}. This 
is in accordance with our rules (section \ref{sec:selection}) and discussed in section \ref{sec:prop_examples}.

\subsection{$\Lambda_b \to \Lambda(1520) (\rho \to \ell^+\ell^-)$ or $1/2^- \to 3/2^+ + 1^-$}
\label{sec:32}
We are going to discuss this example by using the following interaction:
\begin{eqnarray}
\label{eq:H32}
H_{\la_\Lambda\la_\rho} &\propto&  \matel{\rho(q,\la_\rho) \Lambda(p,\la_\Lambda)}{ \Phi(\rho)^\mu \bar s \ga_\mu b}{\Lambda_b} \nonumber \\[0.1cm]
 &=&
\overline \Psi^\mu(p,\la_\Lambda) u(p+q, \la_\Lambda-\la_\rho)  \epsilon_\mu^*(q,\la_\rho) f(q^2)   \;,
\end{eqnarray}
where $\Phi(\rho)$ is an interpolating operator for the $\rho$-meson.
  Here $\la_\rho$ and $\la_\Lambda$ denote polarisation indices, $\Psi_\mu$ is 
a Rarita-Schwinger spin 3/2 object \cite{RS},  $u$ is a Dirac spinor and $f$ is a form factor (irrelevant for our purposes as it evaluates to a constant at the endpoint).  
The decay of the $\rho$-mesons to leptons is not analysed, as it merely serves the possibility of an off-shell $\rho$-meson.
The Rarita-Schwinger $3/2$-spinor is formed out of a Dirac spinor and a spin polarisation tensor 
using CGC:
\begin{equation}
\label{eq:build_RS}
\Psi_\mu(p,\lambda) = \sum_{s = -1/2}^{1/2} C^{\frac32 \frac12 1}_{\lambda s (\lambda-s)} u(p,s) \epsilon_\mu(p,\lambda-s)  \;.
\end{equation}
In order to evaluate \eqref{eq:H32} we use $\bar u(p,\kappa) u(p,\kappa') = \delta_{\kappa\kappa'}$ e.g. \cite{Weinberg:1995mt}
and $\epsilon^*(p,\kappa) \cdot \epsilon^*(q,\kappa')   \stackrel{\eqref{eq:wrong}}{=} - (-)^\kappa \delta_{\bar \kappa \kappa'}$ (where $x \cdot y = x_\mu y^\mu$ throughout), which implies $s = \la_\Lambda- \la_\rho$. Assembling we get:
\begin{equation}
\ha{\la_\Lambda\la_\rho} = - (-)^{\la_\rho}  f C^{\frac32 \frac12 1}_{\la_\Lambda(\la_\Lambda+\bar \la_\rho)   \la_\rho} = \sqrt{2} f C^{ \frac12 \frac32  1}_{ (\la_\Lambda+\bar \la_\rho) \la_\Lambda \bar \la_\rho}\;,
\end{equation}
which corresponds  to Eq. \eqref{eq:geometry} with $a^{(0)} = \sqrt{2}f$. In the second equality we have used the CGC-property $C^{J j_1 j_2}_{M m_1 m_2} = (-)^{j_2+m_2} \sqrt{(2J+1)/(2 j_1+1)}  C^{j_1 J j_2}_{ m_1 M \bar m_2}$ \cite{Hecht}.

Through this concrete example we have aimed to exemplify some of the abstract statements 
made in section \ref{sec:HA}. 
Furthermore we notice that if we had chosen an axial interaction instead of the vector
one then all HAs would have vanished by virtue of  $\bar u(p,\kappa) \ga_5 u(p,\kappa') = 0$. 
The vanishing of the other parity interaction  is in line with the arguments given in section \ref{sec:half-integer} and 
the rule (iii) in section \ref{sec:selection}.

\subsection{NLO power correction: $H_{J=2} \to ZZ$}
\label{sec:NLO-example}

In this section we wish to illustrate the discussion of the NLO power corrections in section \ref{sec:NLO} 
on the example of  Higgs-like spin $2$ particle decaying into two $Z$-bosons. 
The leading order discussion is given in section \ref{sec:ZZ}  under $2 \to 1 + 1$.
We are left with the task of combining the $\bf{2}_{A}, \bf{1}_{B}, \bf{1}_{C}, \bf{1}_{\pI}$ (where the subscripts  
show the association with the particles and the momentum) 
representations into an invariant. This can be done in two different ways since:
\begin{eqnarray}
\label{eq:2inv}
& & ( \mathbf{2_A } \otimes \mathbf{1_{\pI}})_{SU(2)}\otimes (\mathbf{1_B} \otimes\mathbf{1_C})_{SU(2)}  =  \nonumber \\[0.1cm]
& &  (\mathbf{3} \oplus \mathbf{2} \oplus \mathbf{1})_{SU(2)} \otimes (\mathbf{2} \oplus \mathbf{1} \oplus \mathbf{0})_{SU(2)} = 
2 \cdot {\mathbf{0}} \oplus ..
\end{eqnarray}
Note  one of them vanishes. This can be seen from the fact that since 
  $\Delta J + \Omega =1$ is odd, according to rule (iii) section \ref{sec:selection}, 
the amplitude is formed out of a LCT. 
Hence we have to pick the antisymmetric part of the $(\bf{1}_B \otimes \bf{1}_C)$ product 
 which is given by the $\mathbf{1}$ representation. The $\mathbf{2} \otimes \mathbf{2}$ invariant has to vanish.\footnote{\label{foot:two} An example where there are two non-vanishing invariants at NLO is given by $\Lambda_b \to \Lambda \ell \nu$ \cite{HZ21}.
One has  $\mathbf{\frac{1}{2}_A } \otimes   \mathbf{1_B} \otimes
 \mathbf{\frac{1}{2}_{C}} \otimes \mathbf{1_{\pI}}  =  2 \cdot \mathbf{0} \oplus$ 
 and the same applies when the $C$-particle is of spin $\frac{3}{2}$.}
 Hence there is single correction
and this correction can, according to Eqs.~(\ref{eq:scaling2},\ref{eq:2inv}), be computed as 
follows\footnote{\label{foot:detail} To be more concrete we give the example of the covariants formed by 
 the two Clebsch-Gordan products.   The first one is the combination of 
two spin $1$ into a spin $2$ objects and corresponds to $X^{\ga \de} = \epsilon^{\al \be\ga \de} \epsilon^*_\al(p_B) \epsilon^*_\be(p_C)$ 
and the second term is the combination of a spin $2$ object and  a spin $1$ object into a spin two object 
$Y_{\ga\de} = \epsilon_{\{\ga\la}(p_A)q^\la q_{\de\}}$. 
The contracted invariant $X \cdot Y$ will be linear in the velocities as one can verify using 
any specific combinations of non-vanishing helicity combinations.  
In Eq. 19 in \cite{Hopkins12} the above example corresponds to $c_6$. 
The $c_{5,7}$-terms correspond  to   Lorentz invariants  of higher order in the velocities.}
 \begin{equation}
 \label{eq:thisTYPE}
 H_{\la_B \la_C}^{(1)} = a^{(1)}
 C^{211}_{\la_A \la_B \bar \la_C} C^{221}_{\la_A \la_A 0} \;,
 \end{equation}
 where $\ha{\la_B}{\la_C} = \ha{\la_B}{\la_C}^{(0)} + \ha{\la_B}{\la_C}^{(1)}v + {\cal O}(v^2)$ in the notation introduced in \eqref{eq:scalingNLO}.
 When evaluated we get:
 \begin{equation}
 \label{eq:NLOex}
 H_{00} : H_{11} : H_{0\bar 1} : H_{10} : H_{1\bar 1} = 0 : 1 : -\sqrt{3/4} : \sqrt{3/4} : 0 \;.
 \end{equation}
All others can be obtained through $H_{\la_C \la_D} = - H_{\bar \la_C \bar \la_D}$ 
from \eqref{eq:inversion} since $\Delta J  = 1+1-2 =0$ and $\eta = -1$ by rule (ii) in section \ref{sec:selection}.
One may verify from the formulae in Eq.21 \cite{Hopkins12} that \eqref{eq:NLOex} indeed holds with $a^{(0)}_1 \propto c_6$ in their notation.

\subsection{Parity and amplitude properties of example HAs}
\label{sec:prop_examples}

In this section we  illustrate  rules (i), (ii) and (iii) stated in section 
\ref{sec:selection}  for all  eight example endpoint-HAs considered so far.

The parity quantum number of the $Z$-boson is ill-defined as it is a mixture of 
$J^{P} = 1^{-} , 1^{+}$-state even in the case of CP-conservation. Below 
we  simply use $\eta_Z^2 = 1$.\footnote{N.B. of course algebraically 
we would find the same amplitude if say $J_B^{P_B} = 1^-$ and $J_C^{P_C} = 1^+$ 
but then this would imply that $\eta_A$ has got the opposite parity from the one
shown in the table.} The same remark applies to the $W$-boson but rule (i) and (iii) 
imply that only the $1^+$-component couples at the endpoint and thus $\eta_W = 1$ 
is the outcome and strictly speaking an abuse of notation.
 At last we remind the reader that the parity of a Dirac fermion and an anti-Dirac 
fermion are opposite to each other.

\begin{center}
\begin{tabular}{l l   || l  || c c c}
decay &   section &      $\eta = \eta_A \eta_B \eta_C$ \eqref{eq:DJeta} &  $\Delta J$ \eqref{eq:DJeta} & $\Omega$ \eqref{eq:Om}     & amp   \\ \hline
 $0^{\eta_A} \to 1 , 1$  & \ref{sec:ZZ} &  $\eta =  \eta_A  \Rightarrow \eta_A =1$  & even & $0$ &  $g$ \\ 
 $1^{\eta_A} \to 1,  1 $ & idem & $\eta = \eta_A  \Rightarrow \eta_A =1$   &  odd & $0$ & $\epsilon$ \\ 
 $2^{\eta_A} \to 1,  1 $  & idem  &  $\eta = \eta_A  \Rightarrow \eta_A =1$ & even & $0$& $g$  \\ \hline
 $0^{\eta_A} \to 1/2^+  1/2^-$  & \ref{sec:ff}  & $\eta =  -\eta_A  \Rightarrow \eta_A =$-$1$    &  odd & $0$ & $\ga_5$  \\ 
 $1^{\eta_A} \to 1/2^+  1/2^-$ & idem  & $\eta =  -\eta_A  \Rightarrow \eta_A =$-$1$ & even& $0$  &  $\mathbf{1}$ \\ 
 $2^{\eta_A} \to 1/2^+   1/2^{-} $   & idem  & $\eta =  -\eta_A  \Rightarrow \eta_A =1$ &  odd &  $1$ &  $\mathbf{1}$  \\ \hline
 $1/2^+ \to 1/2^+   1^{\eta_W}$ & \ref{sec:12} & $\eta = \eta_W \Rightarrow \eta_W =1$ & odd & $0$ &  $\ga_5$ \\
 $3/2^- \to 1/2^+  1^-$  &  \ref{sec:32} & $\eta=1$  & even & $0$ &  $\mathbf{1}$
 \end{tabular}
 \end{center}
 Above amp is short for amplitude and the abbreviated notation $g,\epsilon,\mathbf{1}$ and 
 $\ga_5$ is explained in the paragraph of rule (iii).
 The reader might verify the statement by  inspecting  the explicit results in the literature and to section \ref{sec:32} for the last example.   The $2 \to 1/2 +  1/2$-case is special as the HA is linear 
in $\pz$ ($\Omega =1$) at the endpoint which imposes $\eta = -1$ by rule (ii).

Note if parity is not conserved the internal parity quantum numbers cease to exist and the states become parity admixtures. 

\section{Brief summary of applications}
\label{sec:applications}

HAs of the $A \to B+C$ type are widely used in particle physics in sequential decays, e.g.  
\cite{Richman:1984gh} for examples, 
where the decay chains do not interact with each other. 
A specific case is given 
by $H \to ZZ^* \to 4 \ell$ where the relations might be of use in determining the Higgs quantum numbers.  The relevant question in practice is whether or not 
the gain in the predictive power, due to the endpoint symmetries, can outweigh 
the loss in phase space and thus statistics.
The selection rules presented in section \ref{sec:selection} 
could potentially be used to test the parity properties (parity even, odd or admixture) 
of the Higgs candidate.

We wish to emphasise that it is important to form ratios 
since the differential rates behave schematically as $d \Gamma \propto \pz\, | \sum \text{HA} |^2  d(\text{angles})$ e.g. \cite{PDG}, 
and therefore vanish at least linearly at the endpoint. 
The number of observables, which are often asymmetries in the context of new physics searches, one can form is vast and depends on the number of  detectable final state particles.

Examples are   the \emph{isotropicity} (appendix \ref{app:isotropic}) of the angular distributions of a decay  $A \to (B \to B_1 B_2) C$ when all helicities summed over.  It is noteworthy that for this to be true the HAs have to be non-zero at the endpoint.
If HA $\propto v^n$ ($n > 0$) then the HA remembers the quantisation axis 
through the scalar products with other vectors. 
The isotropicity has implications for other well studied observables:  
\begin{itemize}
\item[-]
Forward backwards asymmetries in that angle,  $$A_{\rm FB} = \int_{-1}^1  d(\cos \theta_B) \textrm{sign}( \theta_B)  \frac{d \Gamma}{\Gamma d \cos \theta_B}  = 0\;, $$ are zero at the endpoint 
since by definition they are sensitive to odd powers in $\cos(\theta_B)$ only.
\item[-] 
The longitudinal polarisation fraction  corresponding 
to the fraction of $0$-helicity (longitudinal) $B$ and $C$-particle detection. 
In a $(J_A=0)  \to J_B + J_B$ decay 
$$F_L \equiv \frac{\Gamma_L}{\Gamma_L + \Gamma_T} \equiv \frac{|H_{00}|^2}{(\sum_{\la_B=-J_B}^{J_B} |H_{\la_B \bar \la_B}|^2)} 
= \frac{1}{(2 J_B +1)} \;,$$ 
 at the endpoint since 
$|H_{\la_B \bar \la_B} |   \propto  | C^{0J_BJ_B}_{0\la_B \bar \la_B}| = 
|(-1)^{(J_B-\la_B)} (2 J_B+1)^{-1/2}| $ \eqref{eq:geometry} is
independent of  $\la_B$.\footnote{
For $H \to ZZ$ it was for instance noted in \cite{Barger:1993wt}  that the  longitudinal polarisation fraction approaches $F_L = 1/(2\cdot 1 + 1) = 1/3$ at the endpoint.} This results reflects the independence of the non-relativistic limit 
on the spin. It  contrasts with the high energy limit $q_A \gg q_B,q_C$ where the $0$-helicity component dominates, in accordance with the equivalence theorem, and $F_L \to 1$.  This statement is easily verified in the examples treated in section \ref{sec:examples} 
by inspecting the expression in the quoted references.
\end{itemize}

\section{Conclusions and discussion}
\label{sec:conclusions}

In this work we have discussed the simplicity of the $A \to B + C$ HAs at the kinematic endpoint.
Our main results are Eq.~\eqref{eq:geometry} and the refinement in Eqs.~(\ref{eq:scaling},\ref{eq:scaling2}) which relate the HAs to CGCs. 
Parity selection rules and general statements on the structure of the HAs at the 
endpoint have been given in section \ref{sec:selection}.
Our findings are illustrated  in section \ref{sec:examples} with a total of eight examples
including parity selection rules. An outlook on possible applications, including 
$H \to ZZ^* \to 4 \ell$, has been given in the previous section.

We wish to reemphasise that it is the special kinematics which singles out a single 
form factor at the endpoint.  As dynamic objects all the form factors exist at the endpoint; 
and beyond in the sense of analytic continuation and crossed processes as usual.
The independence of the HAs on the spin direction has  got the allure of
a non-relativistic phenomenon as the velocity does indeed approach zero in the limit. 
Power corrections to the asymptotic behaviour were discussed in section \ref{sec:NLO}.
It was found that relative linear corrections (${\cal O}(\pz)$) to the leading order behaviour, which are present in the case of parity violation, can  be accommodated for by CGCs \eqref{eq:scaling2}. Relative corrections of ${\cal O}(\pz^2)$  are of kinematic and dynamic origin. 
The dynamic corrections originate from the form factors themselves. A systematic treatment of these effects, possible within some non-relativistic effective theory and logarithmic quantum corrections, is beyond the scope of this paper.

The application to include further particles, e.g. $A \to B + C+ D$ etc has not been discussed 
in this paper but should be possible by grouping particles together, say $C+D$ to $(CD)$ 
and then proceed as before. Let us give some more detail.
The main complication is that there is not 
a special decay axis anymore. One may single out the direction of flight of the $B$-particle 
in the $A$-restframe, rotate the states $C$ and $D$ with Wigner $D$-matrices 
onto the axis and then proceed to form invariants with CGC.  This essentially reduces 
$A \to B + C+ D$ to $A \to B + (CD)$.
Thus we anticipate a result with two Wigner $D$-matrices and at least two CGC. \\

{\bf Acknowledgements:}  
RZ  acknowledges the support of an advanced STFC fellowship. 
We wish to thank Nick Edwards, Gudrun Hiller, Markus Hopfer and Christos Leonidopoulos for useful 
discussions.

\appendix
\setcounter{equation}{0}
\renewcommand{\theequation}{A.\arabic{equation}}
\numberwithin{equation}{section}

\section{Isotropicity of uniangular distribution}
\label{app:isotropic}

\begin{figure}[ht]
\begin{center}
\includegraphics[width=0.7\textwidth]{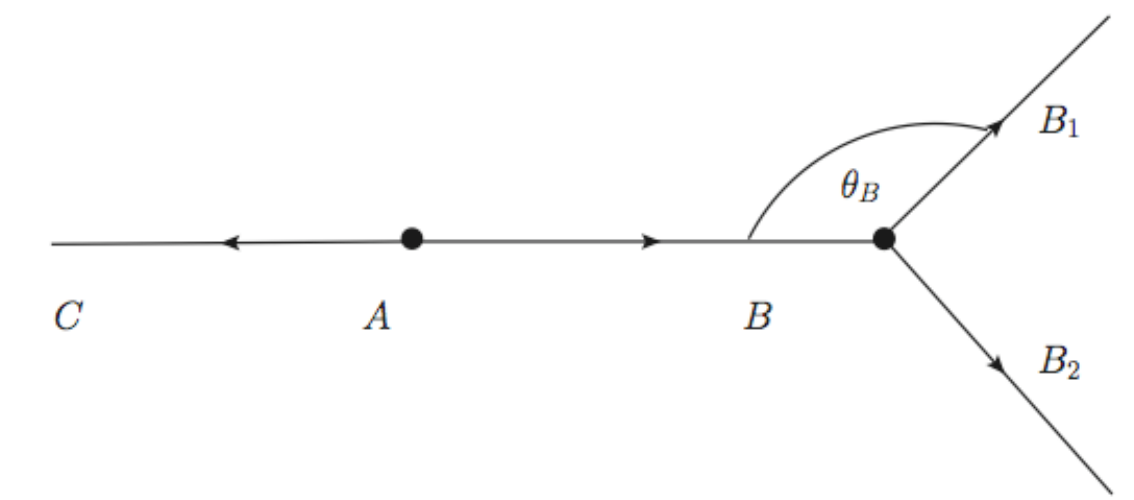}
\end{center}
\caption{\small Decay $A \to  (B \to B_1B_2)C$ in the restframe of particle $A$. 
Whether or not the $C$ particle decays further is immaterial as long as we sum 
over all its final state helicities and integrate over all angles.}
\label{fig:ABC}
\end{figure}

The uniangular distributions of unpolarised initial states 
ought to be \emph{isotropic} (independent of the angle) 
at the kinematic endpoint exhibiting the underlying global $SO(3)$-symmetry. 
An important caveat, to be discussed towards the end of the section, is that the HA must be nonzero at the endpoint for isotropicity to hold.
We consider the  $A \to  (B \to B_1B_2) C$, seen as a two times 
successive $1\to 2$ decays, as depicted in Fig.~\ref{fig:ABC} 
and investigate the uniangular distribution:
\begin{equation}
\frac{d \Gamma}{d \cos \theta_B}(A \to (B \to B_1B_2)C)  = f( \cos \theta_B) \;.
\end{equation}
The physical assumption is that the particle pair $B_1 B_2$ does not interact 
with the $C$-particle in any significant way. 
 Otherwise momentum can be exchanged and invalidate our conclusions.

The amplitude for  $A \to  (B \to B_1B_2) C$ is given by (c.f. \cite{JW59,Dell'Aquila:1985ve} for similar formulae):
\begin{equation}
\label{eq:amp}
{\cal A} \propto  \left(  H^{(A)}_{\la_B \la_C} D^{J_A}_{\la_A , \la_B + \bar \la_C}(\Omega_A) \right) \left(  H^{(B)}_{\la_{B_1} \la_{B_2}} D^{J_B}_{\la_B,\la_{B_1} + \bar \la_{B_2}}(\Omega_B) \right) \;,
\end{equation}
where $D^J_{mm'}(\Omega) \equiv \matel{Jm}{e^{-i \phi_1 J_z} e^{-i \theta J_y} e^{-i \phi_2 J_z}}{Jm'}   $ (with $\Omega = (\phi_1,\theta,\phi_2)$) are the Wigner $D$-functions.  The factors of proportionality that we have dropped in \eqref{eq:amp} are known and have simple dependencies on $J_{A,B,C}$ but are immaterial for our purposes. 
Since $A$ is unpolarised we can choose $\Omega_A = (0,0,0)$ which through $D^J_{mm'}(0) = \delta_{mm'}$ implies helicity conservation on the decay axis:  
$\la_A = \la_B + \bar \la_C$. Further we choose $\Omega_B = (\phi_B , \theta_B,-\phi_B)$.
The decay rate is obtained by squaring the amplitude and summing incoherently over 
all helicity indices of final and initial state particles and coherently over internal helicity indices. Only $\lambda_B$ is of the latter type but does effectively drop out because of
the $\la_A = \la_B + \bar \la_c$ constraint. Whether or not $C$ decays further is immaterial as long as we sum over all helicities of the final state particles and integrate over 
the associated angles. 
The angular distribution is then given by\footnote{For polarised or aligned states one can introduce a density matrix as for instance in \cite{Dell'Aquila:1985ve}. 
Such situations  naturally arise when the HA considered describes an intermediate  
process in a  $A_1 A_2 \to A \to (B \to B_1 B_2) C$-type scattering for example.
N.B. summing over 
initial state helicities is the same as  integrating over all production angles. In the example above this corresponds to the angles of the inverse decay $A \to A_1 A_2$, with respect to the $A \to (B \to B_1 B_2)C$ decay-plane in the $A$-restframe. Formally this is equivalent to taking the density matrix to be the unit matrix.}\begin{equation}
\label{eq:Ga1}
  \frac{d \Gamma}{d \cos \theta_B} \propto \sum_{\la_A,\la_C,\la_{B_1},\la_{B_2}} 
 |H^{(A)}_{\la_B \la_C} |^2 
 |d^{J_C}_{\la_C,\la_{B_1} +\bar \la_{B_2}}(\theta_B)|^2 | H^{(B)}_{\la_{B_1} \la_{B_2}}|^2 \;,
\end{equation}
where  $|d^J_{mm'}(\theta) | = |D^{J}_{mm'}(\Omega)|$ denotes the little Wigner $d$-function 
and is independent of the angles $\phi_{1,2}$.
Taking into account  the endpoint relation \eqref{eq:geometry} Eq.~\eqref{eq:Ga1} can be rewritten as follows
 \begin{equation}
\label{eq:Ga2}
  \frac{d \Gamma}{d \cos \theta_B} \propto \sum_{\la_B,\la_C,\la_{B_1},\la_{B_2}} 
 |C^{J_A J_B J_C}_{(\la_B + \bar \la_C) \la_B \bar \la_C} |^2 
 |d^{J_B}_{\la_B,\la_{B_1} +\bar \la_{B_2}}(\theta_B)|^2 | H^{(B)}_{\la_{B_1} \la_{B_2}}|^2 \;,
\end{equation}
where  $|\la_B-\la_C| \leq J_A$.  
First summing over $\la_B$ using the second relation \eqref{eq:propDC}
and then summing over $\la_C$ using the first relation \eqref{eq:propDC}, 
we may rewrite \eqref{eq:Ga2} as 
\begin{equation}
\label{eq:Ga3}
  \frac{d \Gamma}{d \cos \theta_B} \propto \sum_{\la_{B_1},\la_{B_2}} 
 |H^{(B)}_{\la_{B_1} \la_{B_2}}|^2  =   \text{independent of $\cos(\theta_B)$}\;.
\end{equation}
Thus we have shown that in the case where one sums over all 
initial and final state helicities  
the uniangular distribution in $A \to (B \to B_1B_2)C$ is isotropic; i.e. independent on 
the angle. 

We wish to stress that this does not hold, generally, in the case where the HA vanishes like 
$\pz^\Omega$ with $\Omega > 0$. It is simple matter to use \eqref{eq:scaling2} and 
to construct counter examples. The intuitive reason is that since it scales like
$\pz^\Omega$ the decay has a memory of the decay axis and thus can show preference 
for a certain direction.

\paragraph{Proof of:} 
\begin{equation}
\label{eq:propDC}
 \sum_{m=-J}^J  |d^J_{mm'}|^2 = 1  \;, \quad \sum_{m_1 = -J_1}^{J_1} |C^{JJ_1J_2}_{M m_1 m_2} |^2 = \frac{2J+1}{2J_2+1}  \;,
 \end{equation}
 where $M= (m_1+m_2)$.
The first property follows from 
the definition of the little Wigner $d$-functions:
 $d_{m,m'}^J(\theta) \equiv \matel{J m}{U_R(\theta)}{J m'}$ (e.g. \cite{Hecht}) with 
 $U_R(\theta)=  e^{-\theta J_y}$. One may write,
\begin{eqnarray}
\sum_m  |d^J_{mm' }|^2   &=&  \sum_m \matel{Jm'}{U_R(\theta)}{J m}  \matel{J m}{U^\dagger_R(\theta)}{J m'}  \nonumber \\[0.1cm] 
&=&     \matel{J m'}{U_R(\theta)U^\dagger_R(\theta)  }{J m'} =  \langle Jm' | Jm' \rangle = 1 \;,
\end{eqnarray}
to obtain the result where we have used the fact that $\sum_m |Jm \rangle \langle Jm| = \mathbf{1}_{J}$ is a complete set of states on the Hilbert space of angular momenta $J$.
The second property follows from the so-called  orthogonality relation of the CGC (e.g. \cite{Hecht}) from which one immediately obtains:
\begin{equation}
 \sum_{m_1 m_2} | C^{JJ_1J_2}_{M m_1 m_2}  |^2 = (2 J+1) \;.
\end{equation}
When only the sum over $m_1$, remains the averaging is still sufficient for the result
to be independent of the direction and therefore $m_2$. Thus the sum over $m_2$ 
can be removed 
at the cost of dividing the righthand side by $(2 J_2+1)$ and one therefore arrives at the second formula in \eqref{eq:propDC}.

\section{Polarisation vectors}
\label{app:pol}

We consider the rest frame of the $A$-particle, where $B$ and $C$ are decaying back to back. We denote vectorial ($J=1$) polarisation vector by hatted quantities and parameterise
momenta and $0$-helicity direction as follows
\begin{alignat}{2}
\hat \al(0) &= (0,0,0,1)  \,, \qquad \qquad  \quad &\pA =& (\pA,0,0,0) \nonumber        \\[0.1cm]
\hat \be(0) &= (\pz,0,0,(\pB)_0)/\pB  \,, &\pB =& ((\pB)_0,0,0,\pz)  \nonumber  \\[0.1cm]
\hat \ga(0) &= (-\pz,0,0,(\pC)_0)/\pC  \,,  &\pC =& ((\pC)_0,0,0,-\pz) \;,
\end{alignat}
where $\pB(\pC)_0 \equiv \sqrt{\pB^2  (\pC^2 ) + \pz^2}$ and $\pI = \sqrt{\pI^2}$ with a slight abuse of notation.
The $\pm$-helicity polarisation vectors are written as
\begin{eqnarray}
\hat \al(\pm) = \hat \be(\pm) =  \hat \ga(\mp)  = - (0, \pm 1, i,0)/\sqrt{2} \;, 
\end{eqnarray}
This convention is consistent with the Condon-Shortly phase convention. 
For further important details concerning the relative phase between 
the $\hat \al$, $\hat \be$ and $\hat \ga$ we refer the reader to section \ref{app:second}.
 
Polarisation vectors $\gpol \in \{ \hat \al, \hat \be, \hat \ga \}$
satisfy: 
\begin{equation}
\label{eq:SP}
\gpol(\la_1) \cdot \gpol^*(\la_2) = - \delta_{\la_1 \la_2} \;.
\end{equation}
The minus sign is a remnant of the metric signature $(+,-,-,-)$. 
Throughout this paper $\gpol^*(\la) \equiv (\gpol(\la))^*$.
Furthermore we note that $\gpol^*(\la) = (-1)^\la \gpol(\bar \la)$ and therefore 
\begin{equation}
\label{eq:wrong}
\gpol(\la_1) \cdot \gpol(\la_2) =  -  (-1)^{\la_1} \delta_{\la_1 \bar  \la_2}  \;.
\end{equation}

It is instructive to write a scalar product of two polarisation vectors of non-equal type:
\begin{eqnarray}
\label{eq:gade}
\hat  \be(\la_B) \cdot \hat  \ga^*(\bar \la_C)=  \begin{cases}
-1 & \la_B  =  \la_C = \pm   \\[0.1cm]
(-\pz^2 - \sqrt{\pz^2 + \pB^2} \sqrt{\pz^2 + \pC^2})/( \pB \pC )  & \la_B  = \la_C = 0  \\[0.1cm]
0 & \text{otherwise}\end{cases} \;.  \nonumber  \\
\end{eqnarray}

\paragraph{Endpoint:}
At the endpoint $\pz \to 0$ all polarisation vectors become proportional to 
each other\footnote{N.B. because the quantisation axis of the $C$-particle has been chosen to point into the other direction one has to flip the sign of the helicity index ($\bar \lambda = - \lambda$).}  $\gpol(\la) \equiv \hat \al(\la) = \hat \be(\la) =  \hat \ga(\bar \la)$ and are therefore 
mutually orthogonal,
\begin{equation}
\label{eq:orthogonal}
\hat \al(\la_1) \cdot \hat \be^*(\la_2) = \hat \al(\la_1) \cdot \hat \ga^*(\bar \la_2) = \hat \be(\la_1) \cdot \hat  \ga^*(\bar \la_2)=  - \delta_{\la_1 \la_2} \;.
\end{equation}
We observe that this can also be seen from \eqref{eq:gade} for $\pz \to 0$.

\subsection{Second helicity particle phase convention}
\label{app:second}

It is intuitively clear that the helicity for the 
second spinor moving into the opposite direction as compared 
to the first one is just flipped. On top of that there is some freedom in redefining 
the phase. Let $\chi_a(k, \la)$ ($a = 1,2$) be a two-spinor of the spinorial Lorenzt group 
$SL(2,\mathbf{C})$ then \cite{JW59,Haber:1994pe}
\begin{equation}
\chi_\alpha(-k, \bar \la) = \xi_\lambda \chi_\alpha(k, \la) \;,
\end{equation}
where $-k$ above corresponds to $(k_0,-\vec{k})$.
It is possible to choose the phase $\xi_\lambda = 1$ (Jacob-Wick phase convention \cite{JW59,Haber:1994pe})  which we shall adapt throughout.\footnote{However, if there are multiple channels 
then the phase has to be specified cf. \cite{Marangotto:2019ucc}.} 
Since any polarisation vector can be built from 
the spinor no phases appear for any of them.
In particular for the vector: $  \omega_\mu(-k,\bar \la) =   \omega_\mu(k, \la) $.

\subsection{Higher spin polarisation tensors}
\label{app:higher}
\subsubsection*{Integer spin}
Higher  spin  polarisation tensors of integer spin $J$ denoted by $\gpol(k,\la)_{\mu_1 .. \mu_J}$ 
can be formed out of the $J=1$ polarisation tensor $ \gpol(\la,k)_\mu$  through appropriate Clebsch-Gordan series. This means that the \emph{transversity} property $k \cdot  \gpol(\la,k) = 0$ is inherited:
\begin{equation}
\label{eq:transverse}
k^{\mu_i} \gpol(k,\la)_{\mu_1 .. \mu_i .. \mu_J} =0 \;.
\end{equation}
Two further properties are complete \emph{symmetry} under interchange of indices and \emph{tracelessness}
\begin{equation}
\label{eq:traceless}
g^{\mu_i\mu_j}\gpol(k,\la)_{\mu_1 .. .. \mu_{J}} =0 \;, \quad   1 \leq  i \neq j \leq J \;.  
\end{equation}
Conversely by the symmetry property and the tracelessness are sufficient properties
to find all irreducible representations of $SO(3,1)$ and also of $SL(2,\mathbf{C})$, c.f.
\cite{BK} for  precise statements. 

\subsubsection*{Half-integer spin}
A half integer spin $J = (n + 1/2)$ polarisation tensor can be obtained from a spin $n$ and  
spin $1/2$-polarisation tensor $\left( \mathbf{n} \otimes \mathbf{ 1/2}\right)_{SU(2)} = 1 \cdot (\mathbf{n+1/2}) + ..$, through a single Clebsch-Gordan series.
This is the procedure of Rarita and Schwinger \cite{RS}. The analogue of the tracelessness 
property for the spinorial index is
\begin{equation}
\ga_{\mu_i} \gpol^{\mu_1 .. \mu_i ..\mu_n} = 0 \;,
\end{equation} 
where $\ga_\mu$ is a Dirac matrix. Properties \eqref{eq:transverse} and \eqref{eq:traceless} remain relevant for the Lorentz indices.
  The contraction of the  Dirac index is not shown.


\begin{thebibliography}{199}



\bibitem{JW59}
  M.~Jacob and G.~C.~Wick,
  ``On the general theory of collisions for particles with spin,''
  Annals Phys.\  {\bf 7} (1959) 404
   [Annals Phys.\  {\bf 281} (2000) 774].

\bibitem{Haber:1994pe}
  H.~E.~Haber,
  ``Spin formalism and applications to new physics searches,''
  In *Stanford 1993, Spin structure in high energy processes* 231-272
  [hep-ph/9405376].

\bibitem{Richman:1984gh}
  J.~D.~Richman,
  ``An Experimenter's Guide to the Helicity Formalism,''
  CALT-68-1148.


\bibitem{Korner:1991ph}
  J.~G.~Korner and M.~Kramer,
  ``polarisation effects in exclusive semileptonic Lambda(c) and Lambda(b) charm and bottom baryon decays,''
  Phys.\ Lett.\ B {\bf 275} (1992) 495.

\bibitem{MS}
  A.~D.~Martin and T.~D.~Spearman ...,
  ``Elementary particle theory,''
   North-Holland, Amsterdam, and Elsevier, New York, 1970. xvi, 528 pp


\bibitem{HZ13}
  G.~Hiller and R.~Zwicky,
  ``(A)symmetries of weak decays at and near the kinematic endpoint,''
  JHEP {\bf 1403} (2014) 042
  [arXiv:1312.1923 [hep-ph]].

\bibitem{HZ21}
G.~Hiller and R.~Zwicky,
``Endpoint Relations for Baryons,''
[arXiv:2107.12993 [hep-ph]].



\bibitem{PDG}
P.~A.~Zyla \textit{et al.} [Particle Data Group],
``Review of Particle Physics,''
PTEP \textbf{2020} (2020) no.8, 083C01
doi:10.1093/ptep/ptaa104
Copy to ClipboardDownload


\bibitem{BK}
  I.~L.~Buchbinder and S.~M.~Kuzenko,
  ``Ideas and methods of supersymmetry and supergravity: Or a walk through superspace,''
  Bristol, UK: IOP (1998) 656 p 

  N.~Straumann,
  ``Relativistic quantum theory: An introduction to quantum field theory,''
  Berlin, Germany: Springer (2005) 329 p


\bibitem{Weinberg:1995mt}
  S.~Weinberg,
  ``The Quantum theory of fields. Vol. 1: Foundations,''
  Cambridge, UK: Univ. Pr. (1995) 609 p


\bibitem{Hecht}
K.T.~Hecht in
``Symmetry Properties of Clebsch-Gordan Coefficients,''
Graduate Texts in Contemporary Physics 2000, pp 269-272 Springer New York

\bibitem{DMFV}
  R.~Zwicky and T.~Fischbacher,
  ``On discrete Minimal Flavour Violation,''
  Phys.\ Rev.\ D {\bf 80} (2009) 076009
  [arXiv:0908.4182 [hep-ph]].


\bibitem{Hopkins12}
  S.~Bolognesi, Y.~Gao, A.~V.~Gritsan, K.~Melnikov, M.~Schulze, N.~V.~Tran and A.~Whitbeck,
  ``On the spin and parity of a single-produced resonance at the LHC,''
  Phys.\ Rev.\ D {\bf 86} (2012) 095031
  [arXiv:1208.4018 [hep-ph]].
  


\bibitem{Milleretal}
  D.~J.~Miller, 2, S.~Y.~Choi, B.~Eberle, M.~M.~Muhlleitner and P.~M.~Zerwas,
  ``Measuring the spin of the Higgs boson,''
  Phys.\ Lett.\ B {\bf 505} (2001) 149
  [hep-ph/0102023].

\bibitem{Hopkins10}
  Y.~Gao, A.~V.~Gritsan, Z.~Guo, K.~Melnikov, M.~Schulze and N.~V.~Tran,
  ``Spin determination of single-produced resonances at hadron colliders,''
  Phys.\ Rev.\ D {\bf 81} (2010) 075022
  [arXiv:1001.3396 [hep-ph]].


\bibitem{Dell'Aquila:1985ve}
  J.~R.~Dell'Aquila and C.~A.~Nelson,
  ``$P$ or {CP} Determination by Sequential Decays: 
  ,''
  Phys.\ Rev.\ D {\bf 33} (1986) 80.


\bibitem{Barger:1993wt}
  V.~D.~Barger, K.~-m.~Cheung, A.~Djouadi, B.~A.~Kniehl and P.~M.~Zerwas,
  ``Higgs bosons: Intermediate mass range at e+ e- colliders,''
  Phys.\ Rev.\ D {\bf 49} (1994) 79
  [hep-ph/9306270].

\bibitem{Marangotto:2019ucc}
D.~Marangotto,
Adv. High Energy Phys. \textbf{2020} (2020), 6674595
doi:10.1155/2020/6674595
[arXiv:1911.10025 [hep-ph]].

\bibitem{RS}
  W.~Rarita and J.~Schwinger,
  ``On a theory of particles with half integral spin,''
  Phys.\ Rev.\  {\bf 60} (1941) 61.
  



\end{thebibliography}
\end{document}